\documentclass[sigconf]{acmart}
\AtBeginDocument{%
  \providecommand\BibTeX{{%
    \normalfont B\kern-0.5em{\scshape i\kern-0.25em b}\kern-0.8em\TeX}}}

\copyrightyear{2023}
\acmYear{2023}
\setcopyright{rightsretained}
\acmConference[AIES '23]{AAAI/ACM Conference on AI, Ethics, and Society}{August 8--10, 2023}{Montréal, QC, Canada}
\acmBooktitle{AAAI/ACM Conference on AI, Ethics, and Society (AIES '23), August 8--10, 2023, Montréal, QC, Canada}
\acmDOI{10.1145/3600211.3604673}
\acmISBN{979-8-4007-0231-0/23/08}

\acmConference[ACM AIES 2023] {AAAI/ACM Conference on AI, Ethics, and Society}{August 9-11th 2023}{Montreal, QC}
\usepackage{multirow}
\usepackage{longtable}
\usepackage{booktabs}
\usepackage{graphicx}
\usepackage{wrapfig}
\usepackage{float}
\usepackage{ragged2e}
\usepackage{xcolor}
\usepackage[title]{appendix}
\begin{document}

  \setlength\abovedisplayskip{0pt}
  \setlength\belowdisplayskip{0pt}
%%
%% The "title" command has an optional parameter,
%% allowing the author to define a "short title" to be used in page headers.
\title[Sociotechnical Harms of Algorithmic Systems]{Sociotechnical Harms of Algorithmic Systems: Scoping a Taxonomy for Harm Reduction}
%%
%% The "author" command and its associated commands are used to define
%% the authors and their affiliations.
%% Of note is the shared affiliation of the first two authors, and the
%% "authornote" and "authornotemark" commands
%% used to denote shared contribution to the research.
\author{Renee Shelby}
%\authornotemark[1]
%\email{}
\affiliation{
  \institution{Google Research, JusTech Lab Australian National University}
%  \streetaddress{}
  \city{San Francisco}
  \state{CA}
 \country{USA}
%  \postcode{}
}
\author{Shalaleh Rismani}
%\authornotemark[1]
%\email{}
\affiliation{
  \institution{McGill University}
%  \authornotemark[1]
%  \streetaddress{}
  \city{Montreal}
%  \state{}
  \country{Canada}
%  \postcode{}
}
\author{Kathryn Henne}
%\authornotemark[1]
%\email{}
\affiliation{
  \institution{Australian National University}
%  \streetaddress{}
  \city{Canberra}
%  \state{}
  \country{Australia}
%  \postcode{}
}
\author{AJung Moon}
%\authornotemark[1]
%\email{}
\affiliation{
  \institution{McGill University}
%  \streetaddress{}
  \city{Montreal}
%  \state{}
  \country{Canada}
%  \postcode{}
}

\author{Negar Rostamzadeh}
%\authornotemark[1]
%\email{}
\affiliation{
  \institution{Google Research}
%  \streetaddress{}
  \city{Montreal}
%  \state{}
  \country{Canada}
%  \postcode{}
}
\author{Paul Nicholas}
%\authornotemark[1]
%\email{}
\affiliation{
  \institution{Google}
%  \streetaddress{}
  \city{San Francisco}
  \state{CA}
 \country{USA}
%  \postcode{}
}
\author{N'Mah Yilla-Akbari}
%\authornotemark[1]
%\email{}
\affiliation{
  \institution{Google}
%  \streetaddress{}
  \city{Washington, D.C.}
%  \state{}
  \country{USA}
%  \postcode{}
}
\author{Jess Gallegos}
%\authornotemark[1]
%\email{}
\affiliation{
  \institution{Google Research}
%  \streetaddress{}
  \city{New York City}
  \state{NY}
  \country{USA}
%  \postcode{}
}
\author{Andrew Smart}
%\authornotemark[1]
%\email{}
\affiliation{
  \institution{Google Research}
%  \streetaddress{}
  \city{San Francisco}
  \state{CA}
  \country{USA}
%  \postcode{}
}
\author{Emilio Garcia}
%\authornotemark[1]
%\email{}
\affiliation{
  \institution{Google}
%  \streetaddress{}
  \city{New York City}
  \state{NY}
 \country{USA}
%  \postcode{}
}
\author{Gurleen Virk}
%\authornotemark[1]
%\email{}
\affiliation{
  \institution{Google}
%  \streetaddress{}
  \city{San Diego}
  \state{CA}
 \country{USA}
%  \postcode{}
}

%%
%% By default, the full list of authors will be used in the page
%% headers. Often, this list is too long, and will overlap
%% other information printed in the page headers. This command allows
%% the author to define a more concise list
%% of authors' names for this purpose.
\renewcommand{\shortauthors}{Shelby et al.}

%%
%% The abstract is a short summary of the work to be presented in the
%% article.
\begin{abstract}
Understanding the landscape of potential harms from algorithmic systems enables practitioners to better anticipate consequences of the systems they build. It also supports the prospect of incorporating controls to help minimize harms that emerge from the interplay of technologies and social and cultural dynamics. A growing body of scholarship has identified a wide range of harms across different algorithmic technologies. However, computing research and practitioners lack a high level and synthesized overview of harms from algorithmic systems. Based on a scoping review of computing research (\textit{n}=172), we present an applied taxonomy of sociotechnical harms to support a more systematic surfacing of potential harms in algorithmic systems. The final taxonomy builds on and refers to existing taxonomies, classifications, and terminologies. Five major themes related to sociotechnical harms — representational, allocative, quality-of-service, interpersonal harms, and social system/societal harms — and sub-themes are presented along with a description of these categories. We conclude with a discussion of challenges and opportunities for future research.
\end{abstract}
%%
%% The code below is generated by the tool at http://dl.acm.org/ccs.cfm.
%% Please copy and paste the code instead of the example below.
%%
\begin{CCSXML}
<ccs2012>
   <concept>
       <concept_id>10003456.10003462</concept_id>
       <concept_desc>Social and professional topics~Computing / technology policy</concept_desc>
       <concept_significance>500</concept_significance>
       </concept>
   <concept>
       <concept_id>10002944.10011123.10011130</concept_id>
       <concept_desc>General and reference~Evaluation</concept_desc>
       <concept_significance>500</concept_significance>
       </concept>
 </ccs2012>
\end{CCSXML}

\ccsdesc[500]{Social and professional topics~Computing / technology policy}
\ccsdesc[500]{General and reference~Evaluation}

%%
%% Keywords. The author(s) should pick words that accurately describe
%% the work being presented. Separate the keywords with commas.
\keywords{harms, AI, machine learning, scoping review}

%% A "teaser" image appears between the author and affiliation
%% information and the body of the document, and typically spans the
%% page.
%%\begin{teaserfigure}
%%  \includegraphics[width=\textwidth]{sampleteaser}
%%  \caption{Seattle Mariners at Spring Training, 2010.}
%%  \Description{Enjoying the baseball game from the third-base
%%  seats. Ichiro Suzuki preparing to bat.}
%%  \label{fig:teaser}
%%\end{teaserfigure}

%%
%% This command processes the author and affiliation and title
%% information and builds the first part of the formatted document.
\maketitle

\section{Introduction}
\label{introduction}
Harms from algorithmic systems — that is, the adverse lived experiences resulting from a system’s deployment and operation in the world — occur through the interplay of technical system components and societal power dynamics ~\cite{Green_Viljoen_2020}. This analysis considers how these ``harms (not bounded by the parameters of the technical system)" can ``travel through social systems (e.g., judicial decisions, policy recommendations, interpersonal lived experience, etc.)" \cite[n.p.]{DataSociety2022}. Computing research has traced how marginalized communities — referring to communities that face structural forms of social exclusion ~\cite{Liang_Munson_Kientz_2021} — disproportionately experience sociotechnical harms from algorithmic systems ~\cite{Friedman_Nissenbaum_1996, Graham_2005}. Such experiences include, but are not limited to, the inequitable distribution of resources ~\cite{Crawford_2017}, hierarchical representations of people and communities ~\cite{Noble_2018, Yee_Tantipongpipat_Mishra_2021}, disparate performance based on identity categories ~\cite{BarocasSelbst_2016, mengesha2021don}, and the entrenchment of social and economic inequalities  ~\cite{Benjamin_2019, Eubanks_2018}. In this way, algorithmic systems’ enactment of power dynamics \cite{Orlikowski_2000, henne2021} can function as a minoritizing practice ~\cite{Crooks_Currie_2021} through which unjust social hierarchies are reinforced. 

Practitioners have sought to develop practices that better identify and minimize sociotechnical harms from algorithmic systems (e.g., \cite{Belfield_2020, Cramer_2018, Madaio_2020, Metcalf_Moss_Boyd_2019}).  This includes work to taxonomize harms in HCI on digital safety \cite{ThomasSOK, Scheuerman_Jiang_Fiesler_Brubaker_2021, Agrafiotis_Nurse_Goldsmith_Creese_Upton_2018}, in sociolegal studies on technology-facilitated violence \cite{henry2018technology, Wang_2020}, and canonical responsible ML research on representational, allocative \cite{barocas2017problem}, and quality-of-service harms \cite{blodgett2022responsible} that have significantly shaped the responsible ML field and standards development \cite{Microsoft_2022}. Alongside broader movements towards regulation and standardization ~\cite{Smuha_2021a}, harm reduction practices often draw on fields of auditing, impact assessment, risk management, and safety engineering where a clear understanding of harm is essential ~\cite{Holstein_2019}. Researchers have also developed “ethics methods” ~\cite{Emanuel_Moss_2020} for practitioners to identify and mitigate sociotechnical harms, including statistical assessment ~\cite{Madaio_2020, Metcalf_Moss_Watkins_Singh_Elish_2021}, software toolkits ~\cite{Bird_2020}, and algorithmic impact assessments and audits ~\cite{Raji_Xu_Honigsberg_Ho_2022}, providing notable benefit in how harms are anticipated and identified within algorithmic systems. Existing work on defining, taxonomizing, and evaluating harms from algorithmic systems, however, is vast and disparate, often focusing on particular notions of harm in narrow circumstances. As such, it presents navigational challenges for practitioners seeking to comprehensively evaluate a system for potential harms ~\cite{Raji_Xu_Honigsberg_Ho_2022, Rakova_Yang_Cramer_Chowdhury_2021, Richardson_Garcia-Gathright_Way_Thom_Cramer_2021}, particularly for large generative models that perform different tasks across many use cases. Moreover, the use of different terminologies for describing similar types of harm undermines effective communication across different stakeholder groups working on algorithmic systems ~\cite{Madaio_2020, IEEE2017}.

Recognizing these challenges, we conducted a scoping review \cite{Levac_Colquhoun_OBrien_2010} and reflexive thematic analysis ~\cite{Braun_Clarke_2006} of literature on sociotechnical harms from algorithmic systems, offering a taxonomy to help practitioners and researchers consider them more systematically. A scoping review offers a generative starting place for a landscape harm taxonomy. The purpose of a scoping review is to map the state of a field \cite{arksey2005scoping, Levac_Colquhoun_OBrien_2010}; and here, provides a synthesis of existing articulations of harm, calls attention to forms of harm that may not be well-captured in regulatory frameworks, and reveals gaps and opportunities for future research. As scholarly articulations of harm emerge from different epistemic standpoints, values, and methodologies, this paper pursues the broader question of: How do computing researchers conceptualize harms in algorithmic systems? Three research questions guide this work:
\begin{enumerate}
\label{Research questions}
\item What harms are described in previous research on algorithmic systems? How are these harms framed in terms of their impacts across micro, meso, and macro levels of society? What social dynamics and hierarchies do researchers of algorithmic systems implicate in their descriptions of harms?
\item Where is there conceptual alignment on types of harm from algorithmic systems? What type of organizational structure of harms is suggested by conceptual alignment?
\item How do gaps or absences in research on sociotechnical harms suggest opportunities for future research?
\end{enumerate}

This research contributes to computing scholarship and responsible AI communities, offering:
\begin{itemize}
\item A scoping review of harms, creating an organized snapshot of articulations of computational and contextual harms from algorithmic systems;
\item A reflexive thematic analysis of harms definitions, their impacts to individuals, communities, and social systems, providing a framework for identifying harms when conducting impact and risk assessments on an algorithmic system;
\item Support for interdisciplinary communication by providing terms, definitions, examples of harms, and directions for future work.
\end{itemize}
In what follows, we discuss the sociotechnical character of harms from algorithmic systems and existing harm taxonomies, followed by a description of our methodology (Section \ref{methods}). We then detail the harm taxonomy (Section \ref{findings}), and propose next steps for related work (Section \ref{discussion}). This analysis offers a starting place for practitioners and researchers to reflect on the myriad possible sociotechnical harms from algorithmic systems, to support proactive surfacing and harm reduction.
\section{Background}
\label{background}
\subsection{Sociotechnical Harms} 
Scholars in HCI, machine learning, Science and Technology Studies (STS), and related disciplines have identified various harms from digital technologies (e.g., ~\cite{Angwin_Larson_Mattu_Kirchner_2016, Katzman_Barocas_Blodgett_Laird_Scheuerman_Wallach_2021, Scheuerman_Branham_Hamidi_2018, Scheuerman_Jiang_Fiesler_Brubaker_2021, Smuha_2021b}). This literature underscores harm as a relational outcome of entangled dynamics between design decisions, norms, and power ~\cite{Andalibi_Garcia_2021, Benjamin_2019, DeVito_Walker_Fernandez_2021, Mohamed_Png_Isaac_2020, West_2020}, particularly along intersecting axes of gender \cite{Bivens_Haimson_2016, shelby2021}, race \cite{Noble_2018, henne2021}, and disability \cite{bennett2020, Bennett_Gleason_Scheuerman_Bigham_Guo_To_2021}, among others. Harms from algorithmic systems emerge through the interplay of technical systems and social factors ~\cite{Blodgett_Barocas_Daume_Wallach_2020, Friedman_Nissenbaum_1996} and can encode systemic inequalities ~\cite{Sweeney_2013, May_Wang_Bordia_Bowman_Rudinger_2019, mengesha2021don, blodgett2022responsible}. This duplicity of technology, as Ruha Benjamin ~\cite{Benjamin_2019} explains, is a challenge: algorithms may have beneficial uses, but they often adopt the default norms, and power structures of society.

Recognizing the sociotechnical character of harms from algorithmic systems draws attention to how the development and experience of digital technologies cannot be separated from cultural and social dynamics ~\cite{Alkhatib2021, costanza2020design, qadri2023, prabhakaran2022cultural}. As van Es et al. ~\cite[n.p.]{vanEs_2021} note, “algorithms and code reduce the complexity of the social world into a set of abstract instructions on how to deal with data and inputs coming from a messier reality.” This process involves design decisions predicated on “selection, reduction, and categorization” ~\cite{Bowker_Star_2000} through which technologies come to reflect the values of certain worldviews ~\cite{Bowker_Star_2000, Suresh_Guttag_2021}. Without intentionally designing for equity, algorithmic systems reinforce and amplify social inequalities ~\cite{costanza2020design}.

\subsubsection{Identifying and anticipating harms in practice}
With increased awareness of the need to anticipate harms early in product development ~\cite{Selbst_Boyd_Friedler_Venkatasubramanian_Vertesi_2019}, designers and researchers are central actors in pursuing harm reduction ~\cite{Boyarskaya_2020, Feder_Cooper_Moss_Laufer_Nissenbaum_2022, Green_Viljoen_2020}. Anticipating harms requires considering how technological affordances shape their use and impact ~\cite{Floridi_Strait_2020, Shilton_2015}. It can be done in relation to the technology holistically or with a focus on certain features of the technology and its use by different groups ~\cite{Brey_2012}. This work requires thinking critically about the distribution of benefits and harms of algorithmic systems ~\cite{Birhane_Kalluri_Card_Agnew_Dotan_Bao_2022, Pinch_Bijker_1984} and existing social hierarchies ~\cite{Blodgett_Barocas_Daume_Wallach_2020}. It can be strengthened by bringing in different standpoints and epistemologies, such as feminism ~\cite{DIgnazio_Klein_2021, Noble_2016}, value-based design ~\cite{Ballard_Chappell_Kennedy_2019, Friedman_Hendry_2019, Johnson_2020}, design justice perspectives ~\cite{costanza2020design}, and post-colonial theories ~\cite{Mohamed_Png_Isaac_2020, Taylor_Kukutai_2016}. Importantly, the process requires attending to the constitutive role of social power in producing sociotechnical harms; “designers need to identify and struggle with, alongside the ongoing conversations about biases in data and code, to understand why algorithmic systems tend to become inaccurate, absurd, harmful, and oppressive” ~\cite[p. 2]{Alkhatib2021}. Thus, in anticipating harms, practitioners need to account for computational harms as well as those arising through contextual use ~\cite{Boyarskaya_2020, Passi_Jackson_2018, Weinberg_2022, Saxena2020}.

\subsection{Taxonomies of Harm, Risk, and Failure} 
Structured frameworks can aid practitioners’ anticipation of harms throughout the product lifecycle ~\cite{Madaio_2020, Wong_Boyd_Metcalf_Shilton_2020}. They encourage more rigorous analysis of social and ethical considerations ~\cite{Madaio_2020}, especially when operationalizing principles seems opaque ~\cite{Mittelstadt_2019}. Taxonomizing harms is, however, an exercise in classification, which has potential limitations: taxonomies can draw action to some issues over others, shaping how people navigate and act on information ~\cite{Bowker_Star_2000}. As such, the epistemological choices made in developing harm taxonomies focus attention on certain areas over others ~\cite{Boyarskaya_2020}.

Many existing harm taxonomies address particular domains of use (e.g., ~\cite{Scheuerman_Jiang_Fiesler_Brubaker_2021, Tran_Valecha2019}) and how they are complex assemblages of actors, norms, practices, and technical systems, which can foster individual and collective harm ~\cite{Puar_2020}. Taxonomies have been developed related to online content ~\cite{Banko_Mackeen_Ray, Scheuerman_Jiang_Fiesler_Brubaker_2021, Walker_DeVito_2020}, social media ~\cite{Nova_DeVito_Saha_Rashid_Roy_Turzo_Afrin_Guha, Tran_Valecha2019, Tran_Valecha_Rad_Rao_2020}, users ``at-risk" for experiencing online abuse ~\cite{ThomasSOK, WarfordSOK}, and malicious uses of algorithmic systems \cite{brundage2018malicious}, including cyber attacks ~\cite{Agrafiotis_Nurse_Goldsmith_Creese_Upton_2018} and cyberbullying ~\cite{Ashktorab2018}. Relatedly, they can focus on particular types of harms, such as misinformation \cite{Tran_Valecha_Rad_Rao_2020} or representational harms co-produced through image tagging system inputs and outputs that reinforce social hierarchies ~\cite{Katzman_Barocas_Blodgett_Laird_Scheuerman_Wallach_2021, Wang_Barocas_Laird_Wallach_2022}. While domain-specific taxonomies draw attention to how context informs the emergent nature of harm, they are not easily applicable to a wide range of algorithmic systems. Many systems deploy across contexts, including, for example, ranking and recommendation systems (e.g., search engines or content sorting algorithms on social media platforms) and object detection models (e.g., video surveillance systems, self-driving cars, and accessibility technology).

Another common approach is to orient harm taxonomies around specific algorithmic or model functions (e.g., ~\cite{Ekstrand_Das_Burke_Diaz_2022, Weidinger2022}). Model-focused taxonomies have been developed for large language models ~\cite{Weidinger2022}, image captioning systems ~\cite{Katzman_Barocas_Blodgett_Laird_Scheuerman_Wallach_2021, Wang_Barocas_Laird_Wallach_2022}, and so-called “foundational models,” such as GPT-3 and BERT, which are applied in a wide range of downstream tasks ~\cite{Bommasani_2021}. Organizing harm by model function is highly useful when practitioners’ focus is on a singular model because it draws attention to relevant computational concerns. It does, however, pose limitations to practitioners working downstream on products and features, where multiple model types operate simultaneously, such as in social media, search engines, and content moderation, and where contextual use significantly shapes potential harms. 

Scholars have developed harm taxonomies related to system misbehaviors and failures ~\cite{Bandy2021, Shen_2021}, particularly to aid algorithmic auditing ~\cite{Raji_Kumar_Horowitz_Selbst_2022}. These taxonomies focus on how algorithmic systems are sources of harm (e.g., faulty input/outputs, limited testing, proxy discrimination, and surveillance capitalism) ~\cite{Slaughter_Kopec_Batal}. Bandy ~\cite{Bandy2021} summarizes four problematic behaviors of algorithmic systems — discrimination, distortion, exploitation, and misjudgement. Using such taxonomies focus attention to how specific affordances, training data, and design choices can co-produce harm ~\cite{barocas-hardt-narayanan, Boyarskaya_2020}. Failure-based taxonomies are helpful when practitioners examine potential failure modes of a specific technology, but are often limited in terms of helping to anticipate who or what is harmed. 

In sum, taxonomies can be helpful tools for appreciating and assessing how harms from algorithmic systems are sociotechnical. As they retain social and technical elements, they cannot be remedied by technical fixes ~\cite{Gandy_2010} alone. They require social and cultural change ~\cite{powles2018seductive}. The proposed taxonomy provides a holistic and systematic view of the current discourse on types of sociotechnical harms from algorithmic systems. As the scope of the taxonomy we propose is broad, and many topic- or application- specific taxonomies exist already, we refer to and build on these existing works when appropriate.
\section{Methodology}
\label{methods}
In alignment with prior calls to anticipate computational and contextual harms ~\cite{Boyarskaya_2020, DataSociety2022}, we synthesize insights on harms from computing research to aid anticipation of sociotechnical harms from algorithmic systems. Our findings draw on a scoping review for data collection and a reflexive thematic analysis of computing research on harms.

\subsection{Overview of Methodology}
Our approach followed prior scoping reviews in HCI literature ~\cite{Disch_Fessl_Pammer-Schindler_2022, Ueno_Sawa_Kim_Urakami_Oura_Seaborn_2022}, in alignment with the extension of the PRISMA checklist ~\cite{Liberati_Altman_Tetzlaff_Mulrow_Gotzsche_Ioannidis_Clarke_Devereaux_Kleijnen_Moher_2009}, the PRISMA-ScR (Preferred Reporting Items for Systematic reviews and Meta-Analyses extension for Scoping Reviews) ~\cite{tricco2018prisma}. Scoping studies, as a knowledge synthesis methodology, map existing literature, and “clarify working definitions and conceptual boundaries of a topic or field” ~\cite[p. 141]{peters2015guidance}. They are especially appropriate when distilling and sharing research findings with practitioners ~\cite{tricco2016scoping}, and are suited to identifying evidence on a topic and presenting it visually. Compared to systematic reviews, scoping reviews address a broader range of topics and incorporate different study designs ~\cite{arksey2005scoping}. A scoping review is an effective method for surfacing current ``priorities for research, clarifying concepts and definitions, providing research frameworks or providing background, or contextual information on phenomena or concepts" \cite[p. 2104]{pollock_2021}. We implemented a five-stage scoping review framework ~\cite{Levac_Colquhoun_OBrien_2010, arksey2005scoping}: (1) Identifying research questions; (2) Identifying relevant studies; (3) Study selection; (4) Charting the data; and (5) Collating, summarizing, and reporting results.

\subsubsection{Identify research questions}
To identify the types and range of sociotechnical harms, we developed the three aforementioned research questions (see: Section \ref{Research questions}).

\subsubsection{Identify and select relevant studies}
We then employed multiple strategies to identify relevant resources through different sources: electronic scholarly databases, a citations-based review, and targeted keyword searches in relevant organizations, and conferences. Using the ACM Guide to Computing Literature as the primary search system – which reflects key computing research databases – we developed the following initial set of key concepts to search full text and metadata: “algorithmic harm”, “sociotechnical harm”, “AI harm”, “ML harm”, “data harm”, “harm(s) taxonomy, “allocative harm”, and “representational harm.” Within scoping review methodology, keyword search strategies should be devised to surface relevant literature \cite{arksey2005scoping} and include terms common to the field \cite{hanneke2017scoping}. We included allocative and representational harm as search terms because of their conceptual dominance in machine learning literature since 2017, popularized by responsible ML scholars (e.g., \cite{barocas2017problem, Crawford_2017}), enabling us to surface relevant literature in that sub-field. Iterative searching is a feature of scoping reviews \cite{pollock_2021}. Next, we reviewed each paper, and conducted a citations-based review to surface additional references (e.g., gray literature by nongovernmental organizations (NGOs)) that materially discuss harms, but may not use the specific terminology of the search terms. The citations-based review revealed highly-cited cross-disciplinary scholarship (e.g., articles from sociology, STS). Lastly, we relied on existing knowledge and networks to surface additional sources, including IEEE, NIST, Data and Society, the Aspen Institute, and the AI Incident Database. 

Paper identification started February 2022. The initial search of the ACM database produced a set of 85 research articles (duplicates removed from 118 papers). The citations-based review and targeted keyword searches of NGO and professional organization outputs identified an additional 125 resources. We included articles that described or discussed: (1) algorithmic technologies and (2) harms or adverse impacts from algorithmic systems. We excluded 38 articles that: (1) did not meet the inclusion criteria, and (2) did not have full-text available. In total, 172 articles and frameworks were included in our corpus (see: Appendix Figure 2 and Table 7). 

\subsubsection{Data charting}
We employed a descriptive-analytical method for charting data – a process of “synthesizing and interpreting qualitative data by sifting, charting and sorting material according to key issues and themes" ~\cite[p. 26]{arksey2005scoping}. Two researchers independently charted the following data items extracted through reading of the full text of each source and organized them into a spreadsheet: (1) characteristics of sources: publication year and venue; and (2) description of harm: definition or conceptual framing. Discovery of a new concept or type of harm resulted in a new code, and repeat encounters with existing concepts or harms were documented to reach theoretical saturation – a point at which coding additional papers or resources do not yield additional themes ~\cite{Hennink_Kaiser_Marconi_2017, Saunders_Sim_Kingstone_Baker_Waterfield_Bartlam_Burroughs_Jinks_2018}. It is difficult to know when to stop searching for new sources when conducting a review \cite{hanneke2017scoping}. Thus, relying on scoping reviews' iterative characteristic ~\cite{peters2015guidance}, we used theoretical saturation as a signal to stop sourcing new papers. The entire corpus was coded. Data charting concluded July 2022.  

\subsubsection{Collating and summarizing results}
As collation of themes requires synthesis and qualitative analysis of articles, we used Braun and Clarke’s reflexive thematic analysis ~\cite{Braun_Clarke_2006, Braun_Clarke_2021}, a post-positivist approach to analysis that acknowledges researchers’ standpoints influence data interpretation and encourages self-reflection. Within reflexive thematic analysis, coding is iterative as researchers are immersed in the data. As coding is an evolving, self-reflective process in reflexive thematic analysis, the authors engaged in deep data immersion, interpretation, and discussion, including sharing points of disagreement. First, we thematically sorted definitions of code and looked at the frequency at which harm definitions appeared to begin to identify dominant terms and definitions. Then, we conducted a first line-by-line pass reviewing associated phrasing and terminology of each specific kind of harm. In this initial phase, we identified codes that could be easily condensed. For instance, ‘physical harm’ and ‘physical injury’ were condensed into one code: physical harm. We then began to cluster harms based on the context or domain in which they were mentioned. For example, specific harms describing forms of harassment (e.g., non-consensual sharing of explicit images, or online stalking) were clustered under an initial theme of “hate, harassment, and violence.” There were many conceptual overlaps among harm types; definitions were not always consistent. If there was a dominant term or definition in the cluster that could encompass different sub-types of harm (based on frequency of citation), that term was chosen as the primary category. Notably, we identified and coded more than one type of harm for approximately ~80\% of the articles in the corpus. 

As RQ2 seeks to uncover where there was conceptual alignment across computing sub-fields, initial decisions about harm type and sub-type naming were made after raw coding the entire corpus and discussing emergent themes. From this clustering, and as we iterate from initial codes to final themes, we developed a first version of the harm taxonomy. Three of these harm types — allocative, representational, and quality-of-service — reflect where there was strong definitional and terminological consensus in pioneering responsible ML literature (see: \cite{barocas2017problem, Wang_Barocas_Laird_Wallach_2022, Crawford_2017}). As we iterated from initial codes to final themes we chose to anchor to these canonical harm types in alignment with the RQs. Social system harms and interpersonal harms took shape through the collating and summarizing process. Here, some of the sub-harm types, such as technology-facilitated violence \cite{henry2018technology} and information harms \cite{Wardle_and_Derakhshan_2017} are existing and well-established concepts/terms in different computing sub-fields to which we anchored in alignment with RQ2. See the Appendix for further details on the methodology and descriptive statistics of the corpus.

In scoping reviews, collating and summarizing findings requires researchers to make choices about what they want to prioritize. As the guiding purpose of this research was to develop an applied taxonomy, we prioritized keeping the number of major categories comprehensive yet manageable, envisioning a practitioner with minimal knowledge of harms as the primary user. With the goal of making the taxonomy accessible to practitioners with different disciplinary backgrounds, we repeated this process of clustering and synthesizing three times, refining language and examples of harms to ensure clarity and conceptual cohesion. 

 \begin{figure*}[!t]
    \centering
     \caption{Sociotechnical harms taxonomy overview.}
    \label{fig:Sociotechnical}
    \includegraphics[width=1\textwidth]{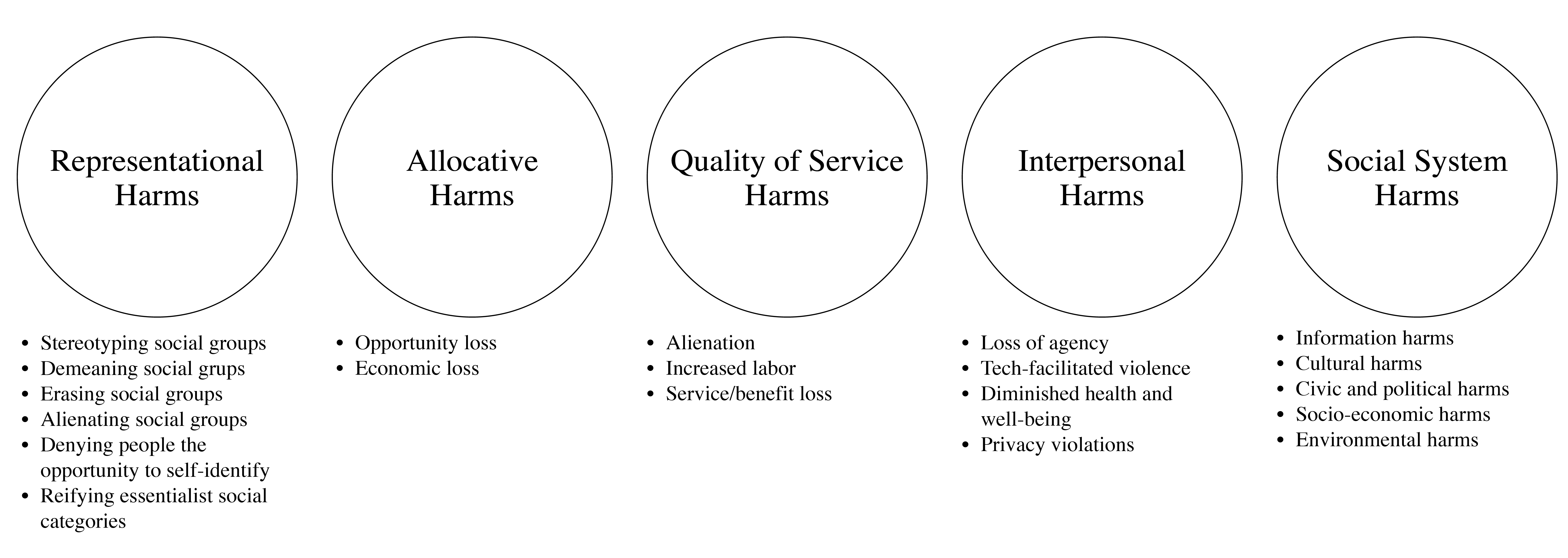}
\end{figure*}

Importantly, while we have aligned to canonical concepts we found definitional variability within and across computing sub-fields, illuminating how understandings of harm are not rigidly fixed and can shift based on sub-field, context of use, technology type, and the evolving state of knowledge. Our descriptions of harm types and sub-types reflect the rich variability that exists in the broader field and is not intended to usurp specific harm definitions that hold specific meaning in different domains (e.g., law, engineering, policy, community work). As such this taxonomy navigates the challenging task of synthesizing multidisciplinary computing research with different priorities and concerns. 

\subsection{Limitations}
In seeking to map how computing researchers conceptualize sociotechnical harms in algorithmic systems, our scoping review focused on academic outlets. The findings are reflective of existing scholarly knowledge for a particular bounded time period. Like all knowledge systems, computing research scholarship is not neutral; it is shaped by various influences, including researcher and institutional priorities, access to resources, thematic conferences, and targeted calls that advance research in specific areas. Focusing on scholarly literature also means vital community-based advocacy addressing the design and use of algorithmic systems is not reflected in the citations. Surveying such work is an important and fruitful direction for future research.

In alignment with feminist standpoint theory ~\cite{harding1991whose, harding1986science}, we prioritize articulations of harm voiced by marginalized communities when possible and foreground these in our descriptions of harm in the taxonomy. We acknowledge articulations of harm described in computing scholarship may derive from work with individuals and communities who describe harms in ways that differ from scholarly discourse. Indeed, there may be many types of harm that are not recognized or articulated in scholarship. As research literatures are always partial and in-progress, the harms described in our taxonomy reflect the partial and in-progress nature of the field. These dynamics are especially relevant to the study of emergent technologies, where individual, collective, and societal impacts of these technologies may be anticipated but not fully known. We also acknowledge the literature reviewed here aligns primarily with Eurocentric worldviews, which undoubtedly shape the descriptions of harms. We are attentive to how these absences likely persist in our taxonomy, having engaged in discussion around how perceived and real gaps in the taxonomy should motivate future research. 

Lastly, the potential benefits of structuring knowledge on sociotechnical harm fosters a paradox: while the taxonomy aids more systematic analysis and minimizes the limitation of relying on the mental models of those at the ``decision-making table," it arguably can hinder practitioners' imagination. These kinds of politics related to knowledge creation have been long critiqued in STS ~\cite{HarawaySituated1988, harding1986science, cetina1999epistemic, fuller1993helen}. Without continued and active critical reflection, this taxonomy — or any structured process the taxonomy is incorporated into — can divert attention away from other possible harms. While we expect understandings of sociotechnical harm will continue to evolve, we encourage those working in this field to retain their critical imagination in considering novel harms. 

\section{Taxonomy of Sociotechnical Harms}
\label{findings}
Our thematic analysis brings together five major types of sociotechnical harms reflective of micro-, meso-, and macro-level impacts of algorithmic systems (see Figure \ref{fig:Sociotechnical} and Appendix Table 6). These categories emphasize (1) how socially constructed beliefs and unjust hierarchies about social groups are reflected in model inputs and outputs (\textit{representational harms}); (2) how these representations shape model decisions and their distribution of resources (\textit{allocative harms}); (3) how choices made to optimize models for particular imagined users result in performance disparities (\textit{quality-of-service harms}); (4) how technological affordances adversely shape relations between people and communities (\textit{interpersonal harms}); and (5) how algorithmic systems impact the emergent properties of social systems, leading to increased inequity and destabilization (\textit{social system}). As our main aim is to provide a cohesive taxonomy for the community, wherever possible, we sought to build on and refer to existing taxonomies, classifications, and terminologies rather than to re-invent new terms. Notably, important concepts such as tech-facilitated violence \cite{henry2018technology, McGlynn_Rackley_2017}, coercive control \cite{Dragiewicz_2018}, disinformation and misinformation \cite{Wardle_and_Derakhshan_2017}, and environmental harms \cite{Bedford_Mann_Foth_Walters_2022} are well-established terms studied in-depth within different computing sub-fields but often alienated from other harms literature. 

In developing a framework that supports more systematic analysis of potential harms in algorithmic systems, we recognize the complex and often concurrent ways harms are experienced. Conceptualizations of harm do not always fit neatly within a compartmentalized structure. Accordingly, there may be gray areas within and across harm categories, and multiple harms may occur in a single use case or system. This taxonomy is thus not prescriptive in its ordering of harms. In suggesting its use as a tool, we encourage considering the multiple dimensions in which harms may play out rather than isolating them. In what follows, we discuss each major harm classification, including sub-types and how they emerge through the interplay of technical components and social dynamics. 

\subsection{Representational Harms: Unjust Hierarchies in Technology Inputs and Outputs}
In our initial coding of the corpus, we located 14 different kinds of representational harms; through the thematic analysis we decided these were analogous to an existing taxonomy on representational harm of image tagging ~\cite{Katzman_Barocas_Blodgett_Laird_Scheuerman_Wallach_2021}. For terminological consistency of the community, we choose to use the same phrasing of the sub-types presented in ~\cite{Katzman_Barocas_Blodgett_Laird_Scheuerman_Wallach_2021}. In this section, we describe representational harms and its sub-types from a broader algorithmic systems perspective. 

Katzman and colleagues describe representational harms as beliefs about different social groups that reproduce unjust societal hierarchies ~\cite{Katzman_Barocas_Blodgett_Laird_Scheuerman_Wallach_2021}. These harms occur when algorithmic systems reinforce the subordination of social groups along the lines of identity, such as disability, gender, race and ethnicity, religion, and sexuality ~\cite{barocas-hardt-narayanan}. Representational harms include instances where certain social groups experience both over- and under-exposure ~\cite{Benjamin_2019}, leading to unequal visibility ~\cite{Yee_Tantipongpipat_Mishra_2021}. Prior work identifies representational harms in many algorithmic systems, including through classifiers ~\cite{Buolamwini_Friedler_Wilson}, natural language processing ~\cite{Blodgett_Barocas_Daume_Wallach_2020}, computer vision ~\cite{Hendricks_Burns_Saenko_Darrell_Rohrbach_2018, Zhao_Wang_Yatskar_Ordonez_Chang_2017,Bennett_Gleason_Scheuerman_Bigham_Guo_To_2021}. Representational harms reflect assumptions that algorithmic systems make about people, culture, and experiences, which perpetuate normative narratives that adversely shape people’s sense of identity and belonging ~\cite{Karizat_2021}. Andalibi and Garcia ~\cite{Andalibi_Garcia_2021} characterize the lived experience of representational harms as \textit{algorithmic symbolic annihilation} through which normative narratives built into technologies become power structures that shape people’s experiences with algorithms. The communities likely to experience these harms are those already experiencing social marginalization. Representational harms thus entrench and exacerbate social stereotypes and patterns of erasure ~\cite{Bengio_Beygelzimer_Crawford_Fromer_Gabriel_Levendowski_Raji_Ranzato_2022}. Specific dimensions of representational harm include stereotyping, demeaning, erasing, and alienating social groups, denying people the opportunity to self-identify, and reifying essentialist social categories (Table \ref{representationalharms}). See Appendix Table 8 for the full list of articles in the corpus that also articulate this harm type.

\subsubsection{Stereotyping social groups} Stereotyping in an algorithmic system refers to how the system's outputs reflect ``beliefs about the characteristics, attributes, and behaviors of members of certain groups....and about how and why certain attributes go together" \cite[p. 240]{Hilton1996}. People marginalized in society face numerous explicit and implicit stereotypes conveyed in various forms of data and coding schema ~\cite{Paullada_Raji_Bender_Denton_Hanna_2021} and design choices ~\cite{Bivens_Haimson_2016} that drive algorithmic systems. Stereotyping often reflects repeated patterns of over- and under-representation — for instance, how gendered beliefs about women’s submissiveness are reflected in digital assistants ~\cite{Cambre_Kulkarni_2019, Sondergaard_Hansen_2018, Wang_2020}. Research identifies narrow stereotypes about masculinity and femininity represented and expressed in natural language processing and computer vision systems, particularly in relation to professions ~\cite{Kay_Matuszek_Munson_2015}, cooking and shopping ~\cite{Zhao_Wang_Yatskar_Ordonez_Chang_2017}, and sport ~\cite{Hendricks_Burns_Saenko_Darrell_Rohrbach_2018}. While computing literature often describes stereotyping along single-axis dimensions of identity \cite{Hendricks_Burns_Saenko_Darrell_Rohrbach_2018, Zhao_Wang_Yatskar_Ordonez_Chang_2017}, an intersectional approach draws attention to how harms play out for people whose lives are shaped by interlocking forms of oppression \cite{Buolamwini_Friedler_Wilson, Wang_Ramaswamy_Russakovsky_2022} -- for example, when a search for the term “unprofessional hairstyles” disproportionately returns images of Black women ~\cite[n.p.]{Alexander_2016}. 

\begin{table}[t]
\centering
\setlength{\tabcolsep}{4.5pt}
\caption{Representational harms}
\label{representationalharms}
\resizebox{\linewidth}{!}{
\begin{tabular}{@{}p{2.70cm}p{7.30cm}}
\toprule
Harm Sub-Type & Example \\ \midrule
Stereotyping \par social groups & “Exclusionary norms {[}in language models{]} can manifest in ‘subtle patterns’ like referring to women doctors as if doctor itself entails not-woman” ~\cite[p. 216]{Weidinger2022}\\  \cline{1-2} 
 Demeaning \par social groups & “A greater percentage of {[}online{]} ads having “arrest” in ad text appeared for Black identifying first names than for white identifying first names in searches” ~\cite[p. 13]{Sweeney_2013}\\  \cline{1-2} 
Erasing \par social groups & “I’m in a lesbian partnership right now and wanting to get married and envisioning a wedding {[}...{]} and I’m so sick of {[}searching for ‘lesbian wedding’ and seeing{]} these straight weddings” ~\cite[p. 13]{DeVos_Dhabalia_Shen_Holstein_Eslami_2022} \\  \cline{1-2} 
 Alienating \par social groups & “{[}Lack of representation{]} further promotes the idea that you don’t belong and perpetuates the sense of alienation” ~\cite[p. 8]{DeVos_Dhabalia_Shen_Holstein_Eslami_2022} \\  \cline{1-2} 
  Denying people \par opportunity to \par self-identify & “It’s definitely frustrating having {[}classifiers{]} get integral parts of my identity wrong. And I find it frustrating that these sorts of apps only tend to recognize two binary genders” ~\cite[p. 12]{Bennett_Gleason_Scheuerman_Bigham_Guo_To_2021} \\  \cline{1-2} 
Reifying essentialist social categories & “{[}Automatic gender recognition{]} aim(s) to capture the morphological sexual differences between male and female faces by comparing their shape differences to a defined face template. We assume that such differences change with the face gender" (quoted in ~\cite[p. 8]{Keyes_2018}) \\ \bottomrule
\end{tabular}}
\end{table}

\subsubsection{Demeaning social groups} In 2013, Latanya Sweeney \cite{Sweeney_2013} drew attention to how algorithmic systems can lead to demeaning treatment of certain social groups. This harm sub-type was also popularized by Safiya Noble in \textit{Algorithms of Oppression} \cite{Noble_2018}. Wang et al. describe demeaning of social groups to occur when they are ``cast as being lower status and less deserving of respect" ~\cite[p. 5]{Wang_Barocas_Laird_Wallach_2022}. This type of representational harm speaks to what sociologist Patricia Hill Collins ~\cite{Collins_1999} calls \textit{controlling images}, referring to discourses, images, and language used to marginalize or oppress a social group. Controlling images include forms of human-animal confusion in image tagging systems ~\cite{Vincent_2018}, which reflect dehumanizing gendered and racialized discourses used to socially exclude and control Black, Indigenous, and other people of color ~\cite{Goff_Eberhardt_Williams_Jackson_2008}. Such controlling images have appeared in ranking and retrieval systems, including reinforcing false perceptions of criminality by displaying ads for bail bond businesses when searching for Black-sounding versus white-sounding names ~\cite{Sweeney_2013}. Similarly, patterns of demeaning imagery have been found in hateful natural language predictions about Muslim people ~\cite{Abid_Farooqi_Zou_2021}, and toxicity and sentiment classifiers that are more likely to classify descriptions or mentions of disabilities ~\cite{Sweeney_2013, Hutchinson_Prabhakaran_Denton_2020} and LGBTQ identities ~\cite{Welbl_2021, Thiago_Marcelo_Gomes_2021} as toxic or negative. As these identities are often weaponized, models struggle with the social nuance and context required to distinguish between hateful and non-hateful speech ~\cite{Welbl_2021}.

\subsubsection{Erasing social groups} Katzman et al. describe that in the context of image tagging, erasing social groups refers to ``when a system fail to recognize—and ... fails to correctly tag people belonging [to] specific social groups or attributes and artifacts that are bound up with the identities of those groups" ~\cite[p. 3-4]{Katzman_Barocas_Blodgett_Laird_Scheuerman_Wallach_2021}. For algorithmic systems more broadly, the erasure of social groups would mean that people, attributes, or artifacts associated with specific social groups are systematically absent or under-represented. Whereas stereotyping reflects systematic patterns of over- and under-representation, erasure captures its extremes. In instances of erasure, certain social groups are not legible to algorithmic systems. Erasure, as Dosono and Semaan ~\cite[p. 1]{Dosono_Semaan_2020} describe, reflects a kind of algorithmic hegemony, as the dominant “system of ideas, practices, and social relations that permeate the institutional and private domains of society” become normalized in sociotechnical systems. Design choices ~\cite{mengesha2021don} and training data ~\cite{Suresh_Guttag_2021} influence which people and experiences are legible to an algorithmic system. Prior work examines erasure in the context of misgendering ~\cite{Dev_Monajatipoor_Ovalle_Subramonian_Phillips_Chang_2021, Keyes_2018}, the systematic erasure of transgender and non-binary people ~\cite{Dev_Monajatipoor_Ovalle_Subramonian_Phillips_Chang_2021, Keyes_2018}, disability and ableism in image descriptions ~\cite{Bennett_Gleason_Scheuerman_Bigham_Guo_To_2021}, and the marginalization of non-Western and underrepresented religious identities in systems ~\cite{Katzman_Barocas_Blodgett_Laird_Scheuerman_Wallach_2021}.   

\subsubsection{Alienating social groups} \label{sec:representative_alienation} Katzman et al. describe alienating as ``when an image tagging system does not acknowledge the relevance of someone's membership in a specific social group to what is depicted in one or more images." \cite[p. 4]{Katzman_Barocas_Blodgett_Laird_Scheuerman_Wallach_2021}. We did not find other work in our corpus that articulate this harm sub-type. This dimension of representational harm diminishes human dignity ~\cite{Mannes_2020} and is especially likely when “a system fails to recognize the injustices suffered by specific social groups” ~\cite[p. 4]{Katzman_Barocas_Blodgett_Laird_Scheuerman_Wallach_2021}. A study of user-elicited identification of harms in image search describes the impacts of such failures as “further promot[ing] the idea that you don’t belong and perpetuat[ing] a sense of alienation” ~\cite[p. 8]{DeVos_Dhabalia_Shen_Holstein_Eslami_2022}. 

\subsubsection{Denying people the opportunity to self-identify} 
Another way algorithmic systems present representational harm is through the complex and non-traditional ways in which humans are represented and classified automatically, and often at the cost of autonomy loss ~\cite{Chancellor_2019, Paullada_Raji_Bender_Denton_Hanna_2021}. As Katzman et al. expresses, in image classification contexts, this means to classify people’s membership without their knowledge or consent ~\cite{Katzman_Barocas_Blodgett_Laird_Scheuerman_Wallach_2021}, such as categorizing someone who identifies as non-binary into a gendered category they do not belong ~\cite{Wang_Ramaswamy_Russakovsky_2022}. This dimension of representational harm reduces autonomy ~\cite{Hanna_Denton_Smart_Smith-Loud_2020} that undermines people’s ability to disclose aspects of their identity on their own terms \cite{corry2021problem}. This loss of autonomy reduces people’s control over data collection, through which data about people, their bodies, and presumptions about their behavior can be extracted into big data flows ~\cite{DIgnazio_Klein_2021}. As classification systems are used across many consequential domains, denying opportunities to self-identify can materially impact marginalized communities, ranging from nonconsensual inclusion in datasets to surveillance and wrongful arrest ~\cite{Bennett_Gleason_Scheuerman_Bigham_Guo_To_2021}. 

\subsubsection{Reifying essentialist social categories} 
Our analysis of the corpus surfaced a type of harm that reinforce social categories as natural, or reinforces perceived classifications of people as truths (see Appendix Table 8). Broadening existing narrower description of this sub-type harm ~\cite{Barocas_Guo_Kamar_Krones_Morris_Vaughan_Wadsworth_Wallach_2021, Katzman_Barocas_Blodgett_Laird_Scheuerman_Wallach_2021}, algorithmic systems that reify essentialist social categories can be understood as when systems that classify a person’s membership in a social group based on narrow, socially constructed criteria that reinforce perceptions of human difference as inherent, static and seemingly natural \cite{Keyes_2018, Devinney_2022}.
Reifying essentialist categories can contribute to “existential harm” in which people are “portrayed in overly reductive terms” ~\cite[p. 162]{Sadowski_Selinger_2014}, often from a Western or Eurocentric perspective ~\cite{Devinney_2022}. When such classification relies on phenotypes, this dimension of representational harm essentializes historically contingent identities ~\cite{Field_Blodgett_Waseem_Tsvetkov_2021} through which classification systems entrench and produce meaning about what they represent ~\cite{Hanna_Denton_Smart_Smith-Loud_2020, Hoffmann_2021}. The harms of reifying social categories are especially likely when ML models or human raters classify a person’s attributes – for instance, their gender, race, or sexual orientation – by making assumptions based on their physical appearance.  

\subsection{Allocative Harms: Inequitable Distribution of Resources}
Allocative harms were first discussed in the ML community by Solon Barocas, Kate Crawford, Aaron Shapiro, and Hanna Wallach (see: \cite{barocas2017problem, Crawford_2017}) and subsequently popularized by Virginia Eubanks in \textit{Automating Inequality} \cite{Eubanks_2018}. Our corpus included 11 thematic codes for allocative harms, which encompass problems arising from how algorithmic decisions are distributed unevenly to different groups of people ~\cite{barocas-hardt-narayanan, Richardson_Gilbert_2021}. These harms occur when a system withholds information, opportunities, or resources ~\cite{barocas-hardt-narayanan} from historically marginalized groups in domains that affect material well-being ~\cite{Microsoft_2022}, such as housing ~\cite{Jane_2018}, employment ~\cite{Singh_Joachims_2018}, social services ~\cite{Bailey_Burkell_Dunn_Gosse_Steeves_2021, Singh_Joachims_2018}, finance ~\cite{Kapania_Siy_Clapper_Sp_Sambasivan_2022}, education ~\cite{Karumbaiah_Brooks_2021}, and healthcare ~\cite{Obermeyer_2019}. Allocative harms “arc towards existing patterns of power” ~\cite[p. 2]{Davis_Williams_Yang_2021} as they entrench material divisions between social groups ~\cite{Zajko_2022}. When occurring in consequential domains, these harms reflect what Mimi Onuhoha ~\cite[n.p.]{Onuhoha} describes as algorithmic violence, in which algorithmic systems “prevent people from meeting their basic needs.” Scholarly literature describes two specific dimensions of allocative harm — opportunity loss and economic loss — reflecting and reinforcing existing social hierarchies along axes of disability, gender, race, or sexuality among others (Table \ref{allocativeharms}). See Appendix Table 9 for the full list of articles in the corpus that also articulate this harm type.

\subsubsection{Opportunity loss} The results of our analysis indicate a sub-type of allocative harms that are conceptually captured by the term ``opportunity loss" presented in a talk by Crawford (cited \cite[n.p.]{barocas-hardt-narayanan}). Opportunity loss occurs when algorithmic systems enable disparate access to information and resources needed to equitably participate in society, including the withholding of housing through targeting ads based on race ~\cite{Angwin_Parris_2016} and social services along lines of class ~\cite{Eubanks_2018}. Researchers contextualize how opportunity loss arises through algorithmic systems and existing patterns of inequality. In relation to housing, for instance, when advertisers target ads based on race and ethnicity, they provide minoritized people fewer options and opportunities to purchase or rent homes ~\cite{Angwin_Parris_2016}. In the employment domain, recommender or ranking systems that match employers and potential candidates may prioritize the resumes of men over other genders ~\cite{Singh_Joachims_2018, vanEs_2021}. Relatedly, these systems may “codify algorithmic segregation” whereby Black candidates are systematically matched to Black-owned businesses and white candidates are systematically matched to white-owned businesses ~\cite[p. 704]{Wu_Mitra_Ma_Diaz_Liu_2022}. In the government or social services domain, screening tools to identify children at-risk for maltreatment can amplify already-existing biases against poor parents ~\cite{Eubanks_2018, Wu_Mitra_Ma_Diaz_Liu_2022}.

\subsubsection{Economic loss} Apart from the loss of opportunity, many articles in our dataset articulate the loss of resources that have negative economic implications (see Appendix Table 9). We refer to these collectively as economic loss. Economic loss is often entwined with opportunity loss, though it relates directly to financial harms ~\cite{cave2019bridging, ONeil_2016} co-produced through algorithmic systems, especially as they relate to lived experiences of poverty and economic inequality. This harm reinforces “feedback loops” between existing socioeconomic inequalities and algorithmic systems ~\cite[p. 14]{Dobbe_Krendl_Gilbert_Mintz_2021}. Researchers recognize economic loss as a harm that intersects with gendered, racialized, and globalized inequalities ~\cite{Avila_2018}. It may arise through different technologies, including demonetization algorithms that parse content titles, metadata, and text, and it may penalize words with multiple meanings ~\cite{Caplan_Gillespie_2020, Duffy_Meisner_2022}, disproportionately impacting queer, trans, and creators of color ~\cite{Duffy_Meisner_2022}. Differential pricing algorithms, where people are systematically shown different prices for the same products, also leads to economic loss ~\cite{Chen_Mislove_Wilson_2016}. These algorithms may be especially sensitive to feedback loops from existing inequities related to education level, income, and race, as these inequalities are likely reflected in the criteria algorithms use to make decisions ~\cite{barocas-hardt-narayanan, Pandey_Caliskan_2021}.

\begin{table}[t]
\centering
\setlength{\tabcolsep}{4.5pt}
\caption{Allocative harms}
\label{allocativeharms}
\resizebox{\linewidth}{!}{
\begin{tabular}{@{}p{2.50cm}p{7.50cm}}
\toprule
Harm Sub-Type & Example \\ \midrule
  Opportunity loss & “Systems…wrongfully deny welfare benefits, kidney transplants, and mortgages to individuals of color as compared to white counterparts” ~\cite[p. 2]{Costanza-Chock_Raji_Buolamwini_2022} \\  \cline{1-2}
 Economic loss & “Language models may generate content that is not strictly in violation of copyright but harms artists by capitalizing on their ideas…this may undermine the profitability of creative or innovative work” ~\cite[p. 221]{Weidinger2022} \\\bottomrule
\end{tabular}}
\end{table}

\subsection{Quality-of-Service Harms: Performance Disparities Based on Identity}
In our initial coding of the corpus, we located 10 different articulations of quality-of-service harms. Performance disparities across different user groups have been widely discussed as a concern in the machine learning community (e.g., ~\cite{Liao_2015, Tatman_2017}). Bird et al. refers to these as quality-of-service harms~\cite{Bird_2020}. These harms occur when algorithmic systems disproportionately underperform for certain groups of people along social categories of difference such as disability, ethnicity, gender identity, and race. DeVries et al. outline that these harms reflect how system training data are optimized for dominant groups ~\cite{DeVries_2019}. Prior work has described how quality-of-service harms are especially likely when system inputs rely on biometric data (e.g., facial features, skin tone, or voice), such as computer vision ~\cite{Buolamwini_Friedler_Wilson, Raji_Gebru_Mitchell_Buolamwini_Lee_Denton_2020}, natural language processing ~\cite{Hutchinson_Prabhakaran_Denton_2020, Liang_Li_Zheng_Lim_Salakhutdinov_Morency_2020, Pruksachatkun_Krishna_Dhamala_Gupta_Chang_2021}, and speech recognition systems ~\cite{koenecke2020racial, mengesha2021don}. Quality-of-service harms are often conceptualized as experiences of directly interacting with an algorithmic system that fails based on identity characteristics, resulting in feelings of alienation, increased labor, and service or benefit loss (see Table \ref{qosharms}). See Appendix Table 10 for the full list of articles in the corpus that also articulate this harm type.

\subsubsection{Alienation} 
Alienation generally refers to ``an individual's feeling of uneasiness or discomfort which reflects [one's] exclusion or self-exclusion from social and cultural participation. It is an expression of non-belonging or non-sharing, an uneasy awareness or perception of unwelcome contrast with others" \cite[p. 758-759]{hadja1961}. Whereas alienation as a form of representational harm diminishes human dignity and the sense of non-belonging (see Section \ref{sec:representative_alienation}), our corpus also surfaced alienation as a quality-of-service harm. In this sub-type, alienation is the specific self-estrangement experienced at the time of technology use, typically surfaced through interaction with systems that under-perform for marginalized individuals ~\cite{mengesha2021don} or reinforce social alienation between humans ~\cite{Bivens_Haimson_2016}. 
In their work on automatic speech recognition systems, Mengesha et al. describe this harm as feelings of annoyance, disappointment, frustration, or anger when interacting with technologies that do not recognize one’s identity characteristics: “Because of my race and location, I tend to speak in a certain way that some voice technology may not comprehend. When I don’t speak in my certain dialect, I come to find out that there is a different result in using voice technology” ~\cite[p. 5]{mengesha2021don}.
Research on trans and queer people’s experiences with voice activated assistants, for instance, describes an awareness of limited representation, noting these technologies “were not designed for trans/or queer people” ~\cite[p. 8]{Rincon_Keyes_Cath_2021}. Similarly, content creators from marginalized communities describe feelings of alienation as they navigate what Duffy and Meisner ~\cite{Duffy_Meisner_2022} refer to as \textit{algorithmic invisibility}, whereby topics important to marginalized communities are rendered invisible by content moderation algorithms.

\subsubsection{Increased labor} In our corpus, certain types of harm surfaced in the form of increased burden (e.g., time spent) or effort required by members of certain social groups to make systems or products work as well for them as others. Research on automatic speech recognition, for instance, has found substantial disparities in word error rates between Black and white speakers (0.35 and 0.19 respectively) ~\cite{koenecke2020racial}. Similar disparities have been found relative to sociolect ~\cite{aksenova-etal-2021-might}, gender ~\cite{aksenova-etal-2021-might, Tatman_2017}, age ~\cite{Liao_2015}, and region ~\cite{aksenova-etal-2021-might}, among others. To correct for these limitations, speakers have to modify their speech to meet system expectations through linguistic accommodation ~\cite{mengesha2021don}.

\subsubsection{Service or benefit loss} Service or benefit loss is the degraded or total loss of benefits of using algorithmic systems with inequitable system performance based on identity ~\cite{mengesha2021don}. Accommodating technology shortcomings limits the potential benefits of technologies. However, when technologies with performance disparities are used in consequential domains — such as in job application videos — degraded service can not only stigmatize users but also lead to other types of harm, such as allocative harms ~\cite{Markl_2022}.

\begin{table}[t]
\centering
\setlength{\tabcolsep}{4.5pt}
\caption{Quality-of-Service harms}
\label{qosharms}
\resizebox{\linewidth}{!}{
\begin{tabular}{@{}p{2.5cm}p{7.5cm}}
\toprule
Harm Sub-Type & Example \\ \midrule
Alienation & “It [voice technology] needs to change because it doesn’t feel inclusive when I have to change how I speak and who I am, just to talk to technology” ~\cite[p. 8]{mengesha2021don}
\\  \cline{1-2} 
Increased labor &  “I modify the way I talk to get a clear and concise response. I feel at times, voice recognition isn’t programmed to understand people when they’re not speaking in a certain way” ~\cite[p. 7]{mengesha2021don} \\  \cline{1-2} 
Service/benefit loss & “It conveyed the opposite message than what I had originally intended, and cost somebody else a lot (of time)” ~\cite[p. 4]{mengesha2021don} \\ \bottomrule
\end{tabular}}
\end{table}

\subsection{Interpersonal Harms: Algorithmic Affordances Adversely Shape Relations}
We initially located 66 thematic codes that we ultimately categorize broadly as interpersonal harms.
Interpersonal harms capture instances when algorithmic systems adversely shape relations between people or communities. As algorithmic systems mediate interactions between people and institutions, interpersonal harms do not necessarily emerge from direct interactions between people, as is the more classic understanding of interpersonal relations, but can emerge through the power dynamics of productionized ML models ~\cite{Henry_Flynn_Powell_2020}. Like other sociotechnical harms, existing power asymmetries and patterns of structural inequality constitutively shape them. They have an intrapersonal element, however, through which people feel a diminished sense of self and agency. Prior work on algorithmic systems describe different types of interpersonal harm, including agency loss, technology-facilitated violence, diminished health and well-being, and privacy violations (Table \ref{interpersonalharm}). See Appendix Table 11 for the full list of articles in the corpus that articulate this harm type.

\subsubsection{Loss of agency or control} Loss of agency occurs when the use ~\cite{Keyes_2018, Malik_Viljanen_Lepinkainen_Alvesalo-Kuusi_Others} or abuse ~\cite{McGlynn_Rackley_2017} of algorithmic systems reduces autonomy. One dimension of agency loss is algorithmic profiling ~\cite{Mann_Matzner_2019}, through which people are subject to social sorting and discriminatory outcomes to access basic services~\cite{Sambasivan_Arnesen_Hutchinson_Doshi_Prabhakaran_2021}. Algorithmic profiling is amplified when there is insufficient ability to contest or remedy the decisions of algorithmic systems ~\cite{Dean_Gilbert_Lambert_Zick_2021, Angwin_Parris_2016, Smith_2021}. As algorithms increasingly curate the flows of information in digital spaces (i.e., recommender systems), Karizat, Delmonaco, Eslami, and Andalibi ~\cite[p. 20]{Karizat_2021} describe how the presentation of content may lead to “algorithmically informed identity change… including [promotion of] harmful person identities (e.g., interests in white supremacy, disordered eating, etc.).” Similarly, for content creators, desire to maintain visibility or prevent shadow banning, may lead to increased conforming of content ~\cite{Thiago_Marcelo_Gomes_2021}.

\subsubsection{Technology-facilitated violence} Technology-facilitated violence occurs when algorithmic features enable use of a system for harassment and violence ~\cite{Afrouz2021, Bailey_Henry_Flynn_2021, Brown_Sanci_Hegarty_2021, Dragiewicz_2018, Henry_Flynn_Powell_2020}, including creation of non-consensual sexual imagery in generative AI. Gender violence scholars have uncovered how algorithmic technologies can become conduits for stalking ~\cite{Freed_Palmer_Minchala_Levy_Ristenpart_Dell_2018, Henry_Flynn_Powell_2020}, online sexual harassment and assault (e.g., sharing images of sexual coercion and violence, sextortion) ~\cite{Brown_Sanci_Hegarty_2021}, and coercive control backed by the threat of violence (e.g., accessing accounts, impersonating a partner, doxxing, sharing sexualized content) ~\cite{Dragiewicz_2018}. For example, abusers may misuse Wi-Fi enabled devices, including locking out and controlling devices to terrorize and harass users ~\cite{Brown_Sanci_Hegarty_2021, PenzeyMoog_Slakoff_2021}, or use technology to generate or share non-consensual sexually explicit images ~\cite{Bailey_Burkell_Dunn_Gosse_Steeves_2021, McGlynn_Rackley_2017}. Beyond gender violence, other facets of technology-facilitated violence, include doxxing ~\cite{Douglas_2016}, trolling ~\cite{Ayodeji_Olamijuwon_Kokomma_Onyemelukwe_Gboyega_2021}, cyberstalking ~\cite{Ayodeji_Olamijuwon_Kokomma_Onyemelukwe_Gboyega_2021}, cyberbullying ~\cite{Ayodeji_Olamijuwon_Kokomma_Onyemelukwe_Gboyega_2021, Guberman_Schmitz_Hemphill_2016, Slonje_Smith_Frisen_2013}, monitoring and control ~\cite{Brown_Sanci_Hegarty_2021}, and online harassment and intimidation ~\cite{Guberman_Schmitz_Hemphill_2016, Scheuerman_Branham_Hamidi_2018, Sheth_Shalin_Kursuncu_2021, Vitak_Chadha_Steiner_Ashktorab_2017}, under the broader banner of online toxicity ~\cite{Guberman_Schmitz_Hemphill_2016, Maity_Chakraborty_Goyal_Mukherjee_2018}. Technology-facilitated violence leads to co-occurring harms, including feelings of distress, fear, and humiliation ~\cite{Brown_Sanci_Hegarty_2021}, while often infringing personal and bodily integrity, dignity, and privacy and inhibiting autonomy and expression ~\cite{McGlynn_Rackley_2017}.

\subsubsection{Diminished health and well-being} 
Algorithmic systems can lead to diminished health and well-being of human users. Our corpus attribute the myriad sources of this harm including from algorithmic behavioral exploitation ~\cite{Bandy2021, Solon_2017}, emotional manipulation ~\cite{Skirpan_Fiesler_2018} whereby algorithmic designs exploit user behavior, safety failures involving algorithms (e.g., collisions) ~\cite{Dean_Gilbert_Lambert_Zick_2021}, and when systems make incorrect health inferences ~\cite{Obermeyer_2019}. They can lead to both physical harms ~\cite{DeVos_Dhabalia_Shen_Holstein_Eslami_2022, Douglas_2016, Mannes_2020, Redden_Brand_2017}  emotional harms, such as distress ~\cite{Weidinger2022}, dignity loss ~\cite{Brown_Sanci_Hegarty_2021, Mannes_2020}, misgendering ~\cite{Keyes_2018}, and reputational harms ~\cite{Moss_2021, Redden_Brand_2017}. Diminished health and well-being may accompany other identified sociotechnical harms – for example, experiences of representational harms, including algorithmic annihilation ~\cite{Andalibi_Garcia_2021} or the internalization of stereotypes may spark other emotional or psychological effects, such as “epistemic doubt” ~\cite{Wang_2020}, which affects overall health. As constructs of well-being are culturally relative ~\cite{Ng_Ho_Wong_Smith_2003}, health ideologies operationalized into machine learning models may “not be relevant, or potentially even harmful, to users living differently to the ways assumed by situated designers” ~\cite[p. 5]{Docherty_Biega_2022}. Thus, builders of algorithmic systems must be attentive to forms of distress falling outside Eurocentric and Western care models ~\cite{Pendse2022}.

\subsubsection{Privacy violation} Privacy violation occurs when algorithmic systems diminish privacy, such as enabling the undesirable flow of private information ~\cite{Ramesh_Kameswaran_Wang_Sambasivan_2022}, instilling the feeling of being watched or surveilled ~\cite{Redden_Brand_2017}, and the collection of data without explicit and informed consent ~\cite{Kapania_Siy_Clapper_Sp_Sambasivan_2022}. These violations have also been framed as “data harms” ~\cite{Monroe-White_2021}, which encompass the adverse effects of data that “impair, injure, or set back a person, entity, or society’s interests” ~\cite[n.p.]{Redden_Brand_2017}. Here, privacy violations may reflect more traditional conceptualizations of privacy attacks or security violations ~\cite{Douglas_2016, Hatzivasilis_2020} and privacy elements beyond what may be protected by regulations or under the traditional purview of a privacy officer ~\cite{Mann_Matzner_2019, Ramesh_Kameswaran_Wang_Sambasivan_2022}. For instance, privacy violations may arise from algorithmic systems making predictive inference beyond what users openly disclose ~\cite{Tufekci_Rit_Adam_Kramer_Guillory_Hancock} or when data collected and algorithmic inferences made about people in one context is applied to another without the person’s knowledge or consent through big data flows ~\cite{Mann_Matzner_2019}, even after those datasets or systems have been deprecated ~\cite{corry2021problem, Ehsan_Singh_Metcalf_Riedl_2022}. Even if those inferences are false (e.g., the incorrect assessment of one’s sexuality), people or systems can act on that information in ways that lead to discrimination and harm ~\cite{Weidinger2022}. Privacy violations may also occur through ubiquitous surveillance, surveillance based on emotional/affective targeting ~\cite{Solon_2017}, or coercive and exploitative data practices ~\cite{Kapania_Siy_Clapper_Sp_Sambasivan_2022}.

\begin{table}[t]
\centering
\setlength{\tabcolsep}{4.5pt}
\caption{Interpersonal harms}
\label{interpersonalharm}
\resizebox{\linewidth}{!}{
\begin{tabular}{@{}p{2.4cm}p{7.6cm}}
\toprule
Harm Sub-Type & Example \\ \midrule
Loss of agency or \par control & “{[}A photo recommender shared a{]} picture of my deceased mother {[}and it{]} just kind of caught me, and I sat there and thought   about different things for a little bit. Then I had to get back to work. But   I was distracted the whole time” ~\cite[p. 8]{Lustig_Konrad_Brubaker_2022} \\ \cline{1-2} 
Technology-facilitated violence & “{[}She{]} broke up with {[}him{]} due to his   controlling behavior. After the break-up, he began to appear where she was…One day, while driving her {[}car{]}, the air conditioner turned off….After a few failed attempts, she figured the unit was broken…After a call with the {[}car’s{]} customer support, she discovered a second person using the {[}car{]} app to connect” ~\cite[p. 650]{PenzeyMoog_Slakoff_2021} \\ \cline{1-2}
%%“A group of people decided to start invading the servers and they made a huge   campaign on the  “aspec” (asexual spectrum) {[}server{]}. They were known to have abused and stalked one user  and they were trying to shit   on the staff  everywhere they could on Tumblr, on Discord.  One was   harassing me via Discord and ... so if they had more personal info about me   they would probably use it in really bad ways.” ~\cite{Scheuerman_Branham_Hamidi_2018}   \\  
Diminished health and well-being & “I was getting ads for maternity clothes. I was   like, ‘Oh please stop.’ …there’s no way to tell your app, ‘I had a   miscarriage. Please stop sending me these updates’” ~\cite[p. 18]{Andalibi_Garcia_2021} \\ \cline{1-2} 
Privacy violations & “{[}Shopping{]} analytics had correctly inferred what he had not known, that his daughter was pregnant.” ~\cite[p. 211]{Tufekci_Rit_Adam_Kramer_Guillory_Hancock} \\ \bottomrule
\end{tabular}}
\end{table}

\subsection{Societal Harms: System Destabilization and Exacerbating Inequalities}
Our corpus included 68 thematic codes that we broadly classify as societal harms. 
Social system or societal harms reflect the adverse macro-level effects of new and reconfigurable algorithmic systems, such as systematizing bias and inequality ~\cite{Eubanks_2018} and accelerating the scale of harm ~\cite{Malik_Viljanen_Lepinkainen_Alvesalo-Kuusi_Others}. Social systems are instantiated through recurrent social practices, shaped by existing and intersecting power dynamics. As Dosono and Semaan ~\cite[p. 2]{Dosono_Semaan_2020} summarize, “people with marginalized identities—those who are pushed to the boundaries of society based on various intersections of their identity such as race and gender—continue to experience oppression, exclusion, and harassment within sociotechnical systems.” Compared to other harm types, social system harms are often indirectly felt and occur downstream; they do not necessarily arise from a single incident or problematic system behavior. Societal harms reflect the “widespread, repetitive or accumulative character” of algorithmic systems in the world ~\cite[p. 10]{Smuha_2021b}, which contribute to institutional exclusions ~\cite{Wang_2020}. Harm to social systems is thus about how algorithmic systems adversely shape the emergent properties ~\cite{Leveson_2016} of social systems ~\cite{Orlikowski_2000}. Prior research outlines such harms in relation to knowledge systems, culture, political and civic harms, socioeconomic systems, and environmental systems (Table \ref{societalharms}). See Appendix Table 12 for articles in the corpus articulating this harm type.

\subsubsection{Information harms} Knowledge systems can be conceived as localized processes through which social knowledge is produced, circulates, and is destabilized. Janzen, Orr, and Terp ~\cite{janzen2022cognitive} use the term information-based harms to capture concerns of misinformation, disinformation, and malinformation. Algorithmic systems, especially generative models and recommender, systems can lead to these information harms. Misinformation refers to the spread of misleading information whether or not there is intention to deceive and disinformation is deliberately false information ~\cite{Neumann_De-Arteaga_Fazelpour_2022, Southwell_Brennen_Paquin_Boudewyns_Zeng_2022, Treen_Williams_ONeill_2020, Wardle_and_Derakhshan_2017}. Malinformation describes “genuine information that is shared with the intent to harm” \cite[p. 2]{janzen2022cognitive}. Information harms are often accompanied by co-occurring impacts, including physical, psychological or emotional, financial, and reputational harms ~\cite{Agrafiotis_Nurse_Goldsmith_Creese_Upton_2018, Tran_Valecha_Rad_Rao_2020, Wardle_Singerman_2021, Wright_Williams_Elizarova_Dahne_Bian_Zhao_Tan_2021}, which scale into broader societal harm. Beyond misinformation, disinformation, and malinformation, knowledge systems may be harmed through “subjugation,” whereby dominant discourses proliferate through algorithmic systems — including in generative language models \cite{Weidinger2022} — and foreclose alternative ways of knowing ~\cite{galaz2021artificial, Sadowski_Selinger_2014, Satra_2020}.

\subsubsection{Cultural harms} Cultures are collectively and dynamically produced ~\cite{Irani_Vertesi_Dourish_Philip_Grinter_2010}. Cultural harm has been described as the development or use of algorithmic systems that affects cultural stability and safety, such as “loss of communication means, loss of cultural property, and harm to social values” ~\cite[p. 30]{Agrafiotis_2016}. As algorithmic technologies can “foreclose alternative ways of understanding the world and restricting imaginations about possible futures” ~\cite[p. 162]{Sadowski_Selinger_2014}, the nature of their harm can encompass adverse cultural impacts such as systemic erasure ~\cite{DeVos_Dhabalia_Shen_Holstein_Eslami_2022}, Eurocentric ideas being exported to Global South ~\cite{Docherty_Biega_2022, Mohamed_Png_Isaac_2020}, harmful cultural beliefs ~\cite{Dobbe_Krendl_Gilbert_Mintz_2021}, such as normalizing a culture of non-consensual sexual activity ~\cite{McGlynn_Rackley_2017}, or proliferating false ideas about cultural groups ~\cite{Dosono_Semaan_2020, Sambasivan_Arnesen_Hutchinson_Doshi_Prabhakaran_2021}.

\subsubsection{Political and civic harms} Political harms emerge when “people are disenfranchised and deprived of appropriate political power and influence” ~\cite[p. 162]{Sadowski_Selinger_2014}. These harms focus on the domain of government, and focus on how algorithmic systems govern through individualized nudges or micro-directives ~\cite{Satra_2020}, that may destabilize governance systems, erode human rights, be used as weapons of war ~\cite{Satra_2021}, and enact surveillant regimes that disproportionately target and harm people of color ~\cite{Katell_Young_Dailey_Herman_Guetler_Tam_Bintz_Raz_Krafft_2020}. More generally, these harms may erode democracy ~\cite{Green_Viljoen_2020}, through election interference or censorship ~\cite{Smuha_2021b}. Moreover, algorithmic systems may exacerbate social inequalities and reduction of civil liberties within legal systems ~\cite{Mannes_2020, Redden_Brand_2017}, such as unreasonable searches ~\cite{Moss_2021}, wrongful arrest ~\cite{Costanza-Chock_Raji_Buolamwini_2022, Costanza-Chock_Raji_Buolamwini_2022, koenecke2020racial}, or court transcription errors ~\cite{koenecke2020racial}. These harms adversely impact how a nation's institutions or services function ~\cite{Agrafiotis_Nurse_Goldsmith_Creese_Upton_2018} and increase societal polarization ~\cite{Smuha_2021b}.

\subsubsection{Macro socio-economic harms} Algorithmic systems can increase “power imbalances in socio-economic relations” at the societal level ~\cite[p. 182]{Agrafiotis_2016, Malik_Viljanen_Lepinkainen_Alvesalo-Kuusi_Others}, including through exacerbating digital divides and entrenching systemic inequalities ~\cite{Irani_Vertesi_Dourish_Philip_Grinter_2010, Wang_Zhao_Van_Kleek_Shadbolt_2022}. The development of algorithmic systems may tap into and foster forms of labor exploitation ~\cite{Docherty_Biega_2022, Mohamed_Png_Isaac_2020}, such as unethical data collection, worsening worker conditions ~\cite{Bengio_Beygelzimer_Crawford_Fromer_Gabriel_Levendowski_Raji_Ranzato_2022}, or lead to technological unemployment ~\cite{cave2019bridging}, such as deskilling or devaluing human labor ~\cite{Png_2022}. For instance, text-to-image models may undermine creative economies ~\cite{Weidinger2022}. While big data flows reshape power within socio-economic systems ~\cite{Mohamed_Png_Isaac_2020, Sambasivan_Arnesen_Hutchinson_Doshi_Prabhakaran_2021}, when algorithmic financial systems fail at scale, these can lead to “flash crashes” and other adverse incidents with widespread impacts ~\cite{Malik_Viljanen_Lepinkainen_Alvesalo-Kuusi_Others}.

\subsubsection{Environmental harms} Environmental harms entail ecological concerns, such as the depletion or contamination of natural resources ~\cite{Bedford_Mann_Foth_Walters_2022, galaz2021machine, Mannes_2020, Mohamed_Png_Isaac_2020, Smuha_2021b, Smuha_2021c, Weidinger2022}, and damage to built environments ~\cite{Mannes_2020}. Ecological harms concern adverse changes to the “ready availability and viability of environmental resources” ~\cite[p. 738]{Metcalf_Moss_Watkins_Singh_Elish_2021} that may occur throughout the lifecycle of digital technologies ~\cite{Png_2022, Welbl_2021} from “cradle (mining) to usage (consumption) to grave (waste)” ~\cite[p. 169]{Bedford_Mann_Foth_Walters_2022}. Similar to other sociotechnical harms, the “benefits and burdens of extractivism are unevenly distributed around the planet” whereby consumption in the economic core are contingent on extraction from the economic periphery ~\cite[p. 170]{Bedford_Mann_Foth_Walters_2022}.

\begin{table}[t]
\centering
\setlength{\tabcolsep}{4.5pt}
\caption{Social system / societal harms}
\label{societalharms}
\resizebox{\linewidth}{!}{
\begin{tabular}{@{}p{2.65cm}p{7.35cm}}
\toprule
Harm Sub-Type & Example \\ \midrule
Information harms & “Users are increasingly exposed to information assembled and presented algorithmically, and many users lack the literacy to comprehend how algorithms influence what they can and cannot see” ~\cite[p. 16]{Dosono_Semaan_2020} \\ \cline{1-2} 
Cultural harms & “{[}An image search for ‘thug’ showing predominantly Black men{]} …It damages all the Black community because if you’re damaging Black men, then you’re hurting Black families” ~\cite[p. 8]{DeVos_Dhabalia_Shen_Holstein_Eslami_2022} \\ \cline{1-2} 
Political and \par civic harms & “Bots, automated programs, are used to spread  computational propaganda. While bots can be used for legitimate functions ... {[}they{]} can be used to spam, harass, silence opponents, ‘give the illusion of   large-scale consensus’, sway votes, defame critics, and spread disinformation campaigns” ~\cite[p. 8]{Redden_Brand_2017} \\ \cline{1-2} 
Macro socio \par economic harms & “Harms associated with the labour and material supply chains of AI technologies, beta testing, and commercial exploitation” ~\cite[p. 1]{Png_2022} \\ \cline{1-2} 
Environmental harms & “The energy cost of training machine learning  models...{[}and{]} harms from intensive water and fuel usage and server farms, consequent chemical and e-waste” ~\cite[p. 5]{Png_2022} \\ \hline
\end{tabular}}
\end{table}
\section{Discussion}
\label{discussion}
In this section, we reflect on the findings of our review, offering a discussion of how the taxonomy may support the anticipation of harms as well as tensions that might arise. We propose directions for continued research on sociotechnical harms to deepen the field’s understanding of conditions that foster such harms.

\subsection{Synthesizing methodological distinctions in studying harms}
Harms from algorithmic systems encompass both computational and contextual harms ~\cite{Boyarskaya_2020, Passi_Jackson_2018, Weinberg_2022, Saxena2020}. Through a scoping review and reflexive thematic analysis, we identify five harm categories, including representational, allocative harms, quality-of-service, interpersonal harms, and social system harms. Our analysis finds authors in the corpus vary widely in their approach to discussing and studying harms — an insight unsurprising given the multidisciplinary focus of the review. Notably, the discussion of harms frequently motivates research on technical aspects of algorithmic systems in the machine learning (ML) literature, but is less often a central analytic or variable. In contrast, research from HCI and related social scientific disciplines often centers harm, but may focus on one harm type, providing rich but narrow insight into that particular harm dimension (e.g., tech-facilitated violence ~\cite{Bailey_Henry_Flynn_2021, Dragiewicz_2018, Henry_Flynn_Powell_2020}). Accordingly, individual scholarly analyses rarely capture the wide scope of harms that can arise from a given algorithmic system, which is a kind of anticipation work required for developing trustworthy and Responsible AI (see: \cite{schwartz2022towards}). 
By synthesizing a distributed body of computing research, this review offers insight into the current discourse and range of harms identified in the literature, which can support Responsible AI efforts to anticipate them in practice. 

\subsection{Towards a shared harms vocabulary with flexible structure} 
As structured frameworks support the work of anticipating harms and other challenges of algorithmic systems ~\cite{Madaio_2020, Wong_Boyd_Metcalf_Shilton_2020}, scoping literature from disparate computing research synthesizes insights that can be useful to practitioners and researchers alike. Here, we offer two findings of note. First, this review reveals dominant areas of concern within computing research on harm, such as the prospect of algorithmic systems exacerbating or scaling existing social inequalities (see Appendix Table 12). While concern for the relational complexities of social inequality is perhaps expected in social science research (where inequality is a core concept), its appearance in ML research might be unexpected given its historic emphasis on statistics. Our findings illuminate inequality as a normative concern of computing research on harm. Given methodological distinctions across computing disciplines, there are opportunities for deeper integration of social science insights into ML approaches to strengthen and enrich understanding of harms from algorithmic systems. There are examples of such work already (e.g., \cite{lee2022, bhatt-etal-2022-contextualizing}).

Second, this review reveals how computing researchers describe harms at varying levels of abstraction. For instance, some authors in the corpus simply refer to “representational harms” in relation to a type of likely harm (e.g., \cite{Davis_Williams_Yang_2021}), while others focus on mid-range articulations. Take, for example, deeper discussions of representational “erasure” (e.g., \cite{Katzman_Barocas_Blodgett_Laird_Scheuerman_Wallach_2021}) and the identification of specific ways in which erasure might manifest in an application (e.g., \cite{Andalibi_Garcia_2021, Dev_Monajatipoor_Ovalle_Subramonian_Phillips_Chang_2021}). The depth in which computing research describes harms is often an artifact of a given study's focus. This observation underscores the need to be able to discuss harms at varying levels of specificity and retain shared understanding. This need is particularly salient in Responsible AI settings where practitioners often interact with different audiences, and may simply refer to ``\textit{harms} broadly, without specifying...harms to who or what" \cite[p. 9]{rismani2022plane}. 

These two findings from the scoping review suggest the resulting taxonomy offers practical benefits, as it cultivates a shared language of harms while accommodating needs to discuss harms with varying specificity. This taxonomy enables what linguists call semantic “entailment” through which clear relationships are established between concepts \cite{korman2018defining}, even though it draws together insights from different epistemological standpoints ~\cite{harding1991whose, harding1986science}. It organizes harm types at cascading levels of conceptual detail, while allowing researchers and practitioners to narrow in on harms under those categories or to further operationalize them at a more granular level. %Thus, we do not envision the proposed taxonomy as a rigid set of categories but as framework of harm that can be expanded and contracted in relation to context and level of abstraction needed. 
%In doing so, the taxonomy can be thought of not as a spectrum, but as an accordion that can move and bend as needed while still maintaining its shape. 
This flexible knowledge structure also supports variable needs that arise when communicating about harms to different audiences. 
For instance, a practitioner working on policy may need higher levels of abstraction to provide generalized guidelines for an application; an engineer benchmarking harms within a model may desire more specificity to operationalize them within an evaluation dataset. 
%The taxonomy supports the different kinds of anticipation work that are critical to harm reduction. With multidisciplinary stakeholders building and governing of algorithmic systems, experts have argued for an adaptive approach that responds to and anticipates emerging sociotechnical risks to manage the inherent uncertainties accompanying innovation ~\cite{whitford2021governance, brass2021adaptive, ulbricht2022algorithmic}. These challenges compound when considering governance decisions are often made against the backdrop of information asymmetries, policy uncertainty, structural power dynamics, and the inevitably of some design errors ~\cite{taeihagh2021assessing}. Different interventions are necessary to address these challenges, and are not limited to formal policy; as algorithmic systems have contributed to the rise of “design-based governance” ~\cite{gritsenko2022algorithmic}, technical assessments and guidelines are especially important. 
A flexible harms taxonomy, such as the approach we suggest here, accommodates the varying needs of practitioners and researchers, while fostering common language to support collective efforts to reduce harms.

\subsection{Navigating tensions between known and emergent harms} 
Encouraging practitioners to reflect on potential harms throughout the product life cycle can help proactively anticipate harms and limit reliance on reactionary responses. Anticipating harms is best conducted when grounded in a use case ~\cite{Emanuel_Moss_2020, Floridi_Strait_2020, Shilton_2015}, because it provides deeper consideration of the domain of use, impacted downstream communities, and technological affordances ~\cite{Boyarskaya_2020}. Authors in the corpus identify how certain harms are inherent to particular ML models (e.g., the relationship between recommender systems and allocative harms), pointing to some stability between certain algorithmic systems and the prospect of harm. For example, without interventions in place, algorithmic systems trained on text corpora from the social world are likely to contain representational harms across some aspect of identity, as text corpora often reflect the inequalities of the social world \cite{caliskan2017semantics}. Even with recognition of the connections between certain ML models and potential harms, a harms taxonomy may still be useful in supporting systematic analyses, such as considering how harms might extend across categories, including how they might inform and contribute to other harms. Consider relationships between allocative and quality-of-service harms: when algorithmic systems fail based on identity, community groups often have to engage in additional labor to correct those failures, revealing how those harms are interrelated in practice. An opportunity thus arises to consider known harms and explore their interdependence ~\cite{dobbe2022}, which may be useful when anticipating harms for novel systems. 

Anticipating harms for novel systems is especially challenging ~\cite{Boyarskaya_2020}: there are more unknown variables, and practitioners may be asked to anticipate harms early in the development or production stage before there are users or even a prototype. When the type of algorithm is novel — for example, code generation or text-to-image models — there may be limited existing empirical harms research to refer to for guidance. The knowledge base offered by prior work on harms provides researchers and practitioners a generative starting point. The taxonomy synthesizes prior literature as an analytic and topic guide that can be interpreted alongside specific contextual features when looking at applied contexts. %Its flexible structure thus accommodates anticipation work in both existing and novel systems. 

\subsection{Towards multidisciplinary proactive and reflexive harm anticipation}
Our review underscores the diverse theoretical foundations and methodologies that the field of computer research employs in relation to sociotechnical harms. The breadth of harms covered in the taxonomy is a first step in supporting practitioners to more systematically reflect on adverse impacts of algorithmic systems. As we appreciate various practitioner standpoints and challenges, we do not offer normative guidance on identifying, evaluating, or controlling for harms. This taxonomy can supplement and enhance existing assessment processes an organization may have in place, providing a starting point for establishing a shared vocabulary on sociotechnical harms. Furthermore, we acknowledge that assessing the overall social impacts of an algorithmic system using only harm as a framing device can miss other negative implications of technology – for instance, inconveniences that arise from exercising the right to data portability or the right to be forgotten ~\cite{bertram2019, bourtoule2021}. 

In making these recommendations, we recognize they maintain both tensions and shortcomings. Given that harm is a broad concept experienced in myriad ways, any taxonomy is limited. Our scoping review, the corpus of which is comprised primarily from academic and gray literature published in English, presents inherent biases and does not necessarily resonate globally. The existing literature privileges Western perspectives ~\cite{Png_2022, Mohamed_Png_Isaac_2020}, leaving a dearth of perspectives from the majority of the world. This has bearing on what experiences become legible as harm. Since completing the scoping review, there have been refinements in how certain harms are conceptualized (e.g. \cite{selbst2022}). It would be a mistake to consider this taxonomy a final, comprehensive list, or solely employ it to quantify the overall degree of harms a system may pose. Rather, it is a synthesis of knowledge that can be built upon and extended. 

Rather than aim to flatten the diversity of methods and strategies to address harm by overprescribing how the taxonomy could be used, we hope various groups can draw on its insights to strengthen their methods. Developing a shared language can accelerate the capacity building of practitioners across organizations – a key objective in presenting these findings. In addition to the taxonomy, we hope to inspire practitioner communities to embrace and advance more systematic harm reduction methodologies, including 1) continued research to study and measure hams prior to launching a product; 2) increasing efforts to prioritize community-driven articulation of harms; and 3) strengthening multidisciplinary approaches. To ensure harm reduction practices and design strategies are comprehensive, future research should investigate understandings of harm and social benefit with communities traditionally marginalized and excluded from technology development ~\cite{Chuanromanee_Metoyer_2021}. 
\section{Conclusion}
\label{conclusion}
Through a scoping review and reflexive thematic analysis of computing research on harms, we offer this taxonomy of sociotechnical harms as an initial guide to support practitioners and researchers in addressing a range of adverse impacts informed by algorithmic systems. We expect and hope it will evolve as research and engagement progress, particularly in terms of participatory and community-driven research methodologies (e.g., ~\cite{DeVos_Dhabalia_Shen_Holstein_Eslami_2022, Dev_Monajatipoor_Ovalle_Subramonian_Phillips_Chang_2021, Friedman_Hurley_Howe_Nissenbaum_Felten_2002}). As a synthesis of computing research, this taxonomy offers a measure to assess areas and directions for future scholarly and practitioner discussions and research. It reveals there is greater consensus and depth of work in investigating particular harms, such as representational and allocative harms, as well as gaps where the range of possible harms is likely under articulated. Our interest in scoping sociotechnical harms remains motivated by cultivating methods to reduce their likelihood in algorithmic systems. It is our stance that developing a richer understanding of harms creates more generative paths towards harm reduction for all. %Ultimately, we hope taking stock of already-identified sociotechnical harms illuminates and inspires harm-reduction approaches that cross-cut harm types.

%%%\section*{Acknowledgements} 
\begin{acks}
The authors would like to thank Remi Denton, Fernando Diaz, Mark D\'iaz, Gabriela Erickson, Megan Ma, Michael Madaio, Amanda McCroskery, Philip Parham, Bogdana Rakova, Jamila Smith-Loud, Allison Woodruff, Ben Zevenbergen, Lauren Wilcox, and the anonymous reviewers for their comments that contributed to the paper's development. We are also grateful to Solon Barocas, Hanna Wallach, Su Lin Blodgett, and Hal Daum\'e III for helping us to improve the presentation and quality of our work.
\end{acks}

\bibliographystyle{ACM-Reference-Format}
\bibliography{sample-base}

%%
%% If your work has an appendix, this is the place to put it.
%\begin{appendices}
%\input{sec}
%\end{appendices}

\end{document}